\documentclass[preprint2]{aastex61}

\usepackage{natbib}
\newcommand{\kepler}{\textit{Kepler}}
\newcommand{\rprime}{$R^\prime_{HK}$}
\usepackage{amsmath}
\usepackage{hyperref}

\shorttitle{Chromospheric activity of HAT-P-11}
\shortauthors{Morris et al.}

\begin{document}

\title{Chromospheric Activity of HAT-P-11: an Unusually Active Planet-Hosting K Star}

\author{Brett M. Morris}
\affiliation{Astronomy Department, University of Washington, Seattle, WA 98195, USA}
                 
\author{Suzanne L. Hawley}
\affiliation{Astronomy Department, University of Washington, Seattle, WA 98195, USA}

\author{Leslie Hebb}
\affiliation{Physics Department, Hobart and William Smith Colleges, 
                 Geneva, NY 14456, USA}
                 
\author{Charli Sakari}
\affiliation{Astronomy Department, University of Washington, Seattle, WA 98195, USA}

\author{James. R. A. Davenport}
\affiliation{Department of Physics \& Astronomy, Western Washington University, 516 High St., Bellingham, WA 98225, USA}
\affiliation{NSF Astronomy and Astrophysics Postdoctoral Fellow}

\author{Howard Isaacson}
\affiliation{Department of Astronomy, UC Berkeley, Berkeley, CA 94720, USA}

\author{Andrew W. Howard}
\affiliation{Department of Astrophysics, California Institute of Technology, MC 249-17, Pasadena, CA 91125, USA}

\author{Benjamin T. Montet}
\affiliation{Department of Astronomy and Astrophysics, University of Chicago, 5640 S. Ellis Avenue, Chicago, IL 60637, USA}
\affiliation{NASA Sagan Fellow}

\author{Eric Agol}
\affiliation{Astronomy Department, University of Washington, Seattle, WA 98195, USA}

\email{bmmorris@uw.edu}

\begin{abstract}
\kepler\ photometry of the hot Neptune host star HAT-P-11 suggests that its spot latitude distribution is comparable to the Sun's near solar maximum. We search for evidence of an activity cycle in the CaII H \& K chromospheric emission $S$-index with archival Keck/HIRES spectra and observations from the echelle spectrograph on the ARC 3.5 m Telescope at APO. The chromospheric emission of HAT-P-11 is consistent with a $\gtrsim 10$ year activity cycle, which plateaued near maximum during the \kepler\ mission. In the cycle that we observed, the star seemed to spend more time near active maximum than minimum. We compare the $\log R^\prime_{HK}$ normalized chromospheric emission index of HAT-P-11 with other stars. HAT-P-11 has unusually strong chromospheric emission compared to planet-hosting stars of similar effective temperature and rotation period, perhaps due to tides raised by its planet.
\end{abstract}


\section{Introduction}

The K4 dwarf HAT-P-11 bears many spots on its surface, despite its 29 day rotation period. \citet{Morris2017} used \kepler\ photometry to measure spot sizes and positions during transits of its hot Neptune with a nearly polar orbit. They showed that the spots cover $\sim$100x more surface area on HAT-P-11 than sunspots cover on the Sun, and that the spots cluster into active latitudes similar to the solar active latitudes near activity maximum. The analysis by \citet{Morris2017} was limited in time to the four years of \kepler\ photometry, preventing the detection of activity evolution throughout the stellar activity cycle. The transit photometry also restricted their analysis to spots within the transit chord, which covers only 6\% of the observer-facing stellar hemisphere per transit. 

High resolution spectroscopy is a complementary means of quantifying the disk-integrated activity of HAT-P-11 on long timescales. Emission in the cores of the CaII H \& K features is generated in the chromosphere over active regions. CaII H \& K emission is typically measured as the flux in the emission features normalized by two pseudocontinuum regions on either side of the absorption features, called the $S$-index. The $S$ and related \rprime\ indices of solar-like stars has been studied over many years and across spectral types and ages \citep{Wilson1978, Noyes1984, Duncan1991, Baliunas1995}. CaII emission generally declines with age \citep{Skumanich1972} -- with a large intrinsic variance driven by the rotation of active regions into and out of view, and stellar activity cycles \citep[see review by][]{Hall2008}.

Here we measure the $S$-index of HAT-P-11, and put the activity of HAT-P-11 into context among stars of similar spectral type and age by calculating \rprime. In Section~\ref{sec:sindex} we present spectroscopy from Keck/HIRES and Apache Point Observatory (APO) ARC 3.5 m Telescope from 2008-2017 and find evidence for an activity cycle. We compare the activity cycle observed in spectroscopy with the total stellar brightness in the \kepler\ band from full-frame image photometry in Section~\ref{sec:ffi}. In Section~\ref{sec:cks} we investigate whether or not the abundant activity of HAT-P-11 is ``normal'' among stars of similar age and spectral type.

\section{Observations of the $S$-index} \label{sec:sindex}

We can search for the activity cycle of HAT-P-11 if we combine observations from Keck/HIRES and recent observations from the Astrophysical Research Consortium (ARC) 3.5 m Telescope at APO. However, the definition of the $S$-index has the unfortunate property that its value varies from one instrument to the next for the same intrinsic flux. Therefore $S$-index measurements must be calibrated against the stars observed in the Mount Wilson Observatory (MWO) sample before spectra from different observatories can be compared. In Section~\ref{sec:s_apo} we calibrate the linear correction for the $S$-index measured by the ARC Echelle Spectrograph (ARCES) at APO, and in Section~\ref{sec:s_h11} we combine the recent APO spectra with archival Keck/HIRES spectra to span nine years of observations.

\subsection{Calibrating the $S$-index for APO} \label{sec:s_apo}

We reduce the raw ARCES spectra with \texttt{IRAF} methods to subtract biases, remove cosmic rays, normalize by the flat field, and do the wavelength calibration with exposures of a thorium-argon lamp\footnote{An ARCES data reduction manual by J. Thorburn is available at \url{http://astronomy.nmsu.edu:8000/apo-wiki/attachment/wiki/ARCES/Thorburn_ARCES_manual.pdf}}. We fit the spectrum of an early-type star with a high-order polynomial to measure the blaze function, and we divide the spectra of HAT-P-11 and the MWO stars by the polynomial fit to normalize each spectral order.

Next the normalized spectra must be shifted in wavelength into the rest-frame by removing their radial velocities. We remove the radial velocity by maximizing the cross-correlation of the ARCES spectra with PHOENIX model atmosphere spectra \citep{Husser2013}.

\begin{figure}
\centering
\includegraphics[scale=0.4]{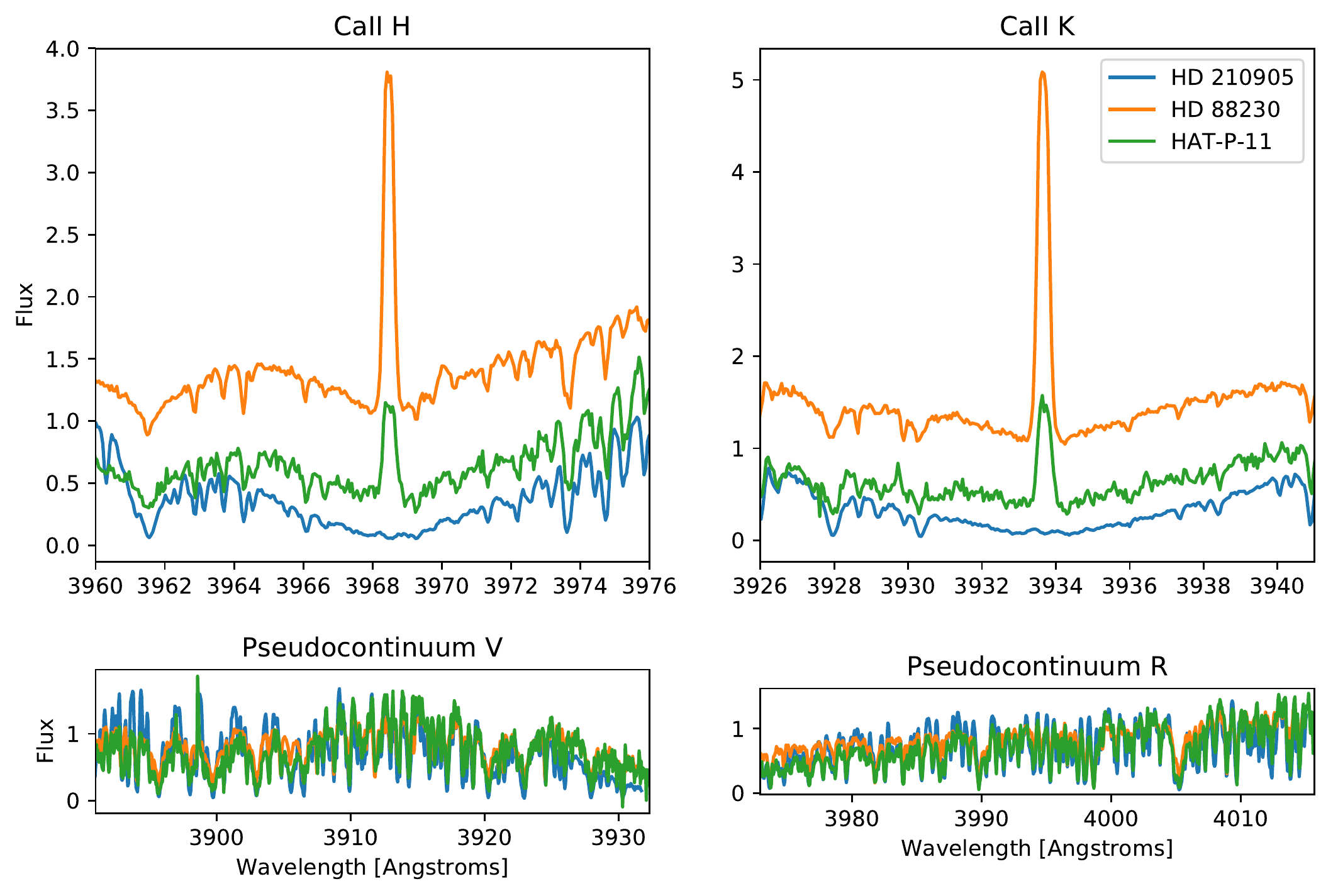}
\caption{Spectra of two CaII H \& K calibration targets on the high and low activity ends, and HAT-P-11 in the middle.}
\label{fig:examplespectra}
\end{figure}

To calibrate the ARCES spectra, we follow the calibration procedure developed in \citet{Isaacson2010} for HIRES. We collect 51 spectra of 30 stars in the \citet{Duncan1991} MWO sample with the ARC 3.5 m Telescope at APO and the ARCES spectrograph, including 22 K stars, 7 G stars and one M star -- see Figure~\ref{fig:examplespectra} for example spectra. 

We measure the $S$-index for these stars by taking the sum of the flux in the cores of the $H$ and $K$ features at 3968.47 \AA\ and 3933.66 \AA, weighted with a triangular weighting function with FWHM=$1.09$\AA. We normalize the weighted emission by the flux in pseudocontinuum regions $R$ and $V$, which are 20 \AA-wide bands centered on 3900 and 4000 \AA, respectively. Then $S$ on the MWO-calibrated scale is 
\begin{eqnarray}
S_{APO} &=& \frac{a~H + b~K}{c~R + d~V} \\
S_{MWO} &=& C_1 S_{APO} + C_2, \label{eqn:s_ind}
\end{eqnarray}
where $a,~b,~c,~d, ~C_1$ and $C_2$ are parameters that can be tuned to make ARCES $S$-indices match the scale of $S_{MWO}$ \citep{Duncan1991}. Following the example of \citet{Isaacson2010}, we chose values of $a,b,c,d$ so that $S$ has roughly equal flux contributions from the $H$ and $K$ emission lines, and roughly equal flux from the $R$ and $V$ psuedocontinuum regions in the APO spectra. Thus we set $a = c = 1$, and we choose $b=2$ and $d=1$, so that $H \sim b~K$ and $K \sim d~V$. 

\begin{figure}
\begin{center}
\includegraphics[scale=0.6]{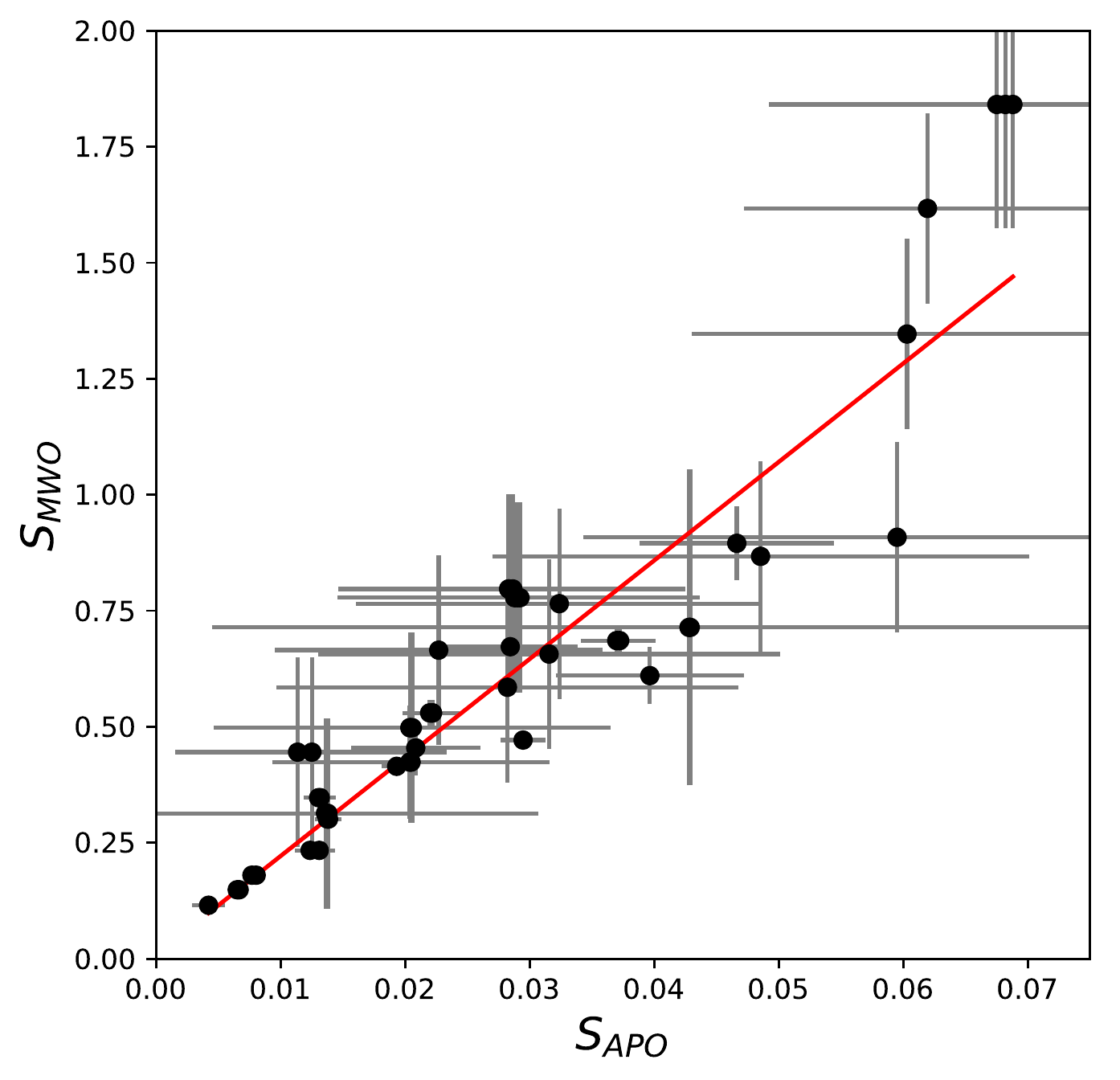}
\caption{We calibrate the $S$-index measured with the APO echelle spectra against previous measurements from Mount Wilson Observatory (MWO) \citep{Duncan1991}. The large uncertainties in the MWO measurements are a result of the intrinsic activity variation of each star, and the uncertainties of the APO observations correspond to the measurement uncertainties.}
\end{center}
\label{fig:calib}
\end{figure}

Since $S$ varies over time for each star in the sample, the linear correlation between the $S_{APO}$ and $S_{MWO}$ will have some intrinsic spread. To incorporate this into our model, we adopt the $\left< S \right>$ and the standard deviation of $S$ from \citet{Duncan1991} as the measurement and uncertainty of the MWO values. We solve for the constants $C_1$ and $C_2$ and their uncertainties with Markov Chain Monte Carlo (MCMC) \citep{Goodman2010, Foreman-Mackey2013}; the results are shown in Figure~\ref{fig:calib}. We find $C_1 = 21.26_{-0.83}^{+0.99}$ and $C_2 = 0.009_{-0.009}^{+0.011}$. The $S$-indices for each target are enumerated in Table~\ref{tab:cals}. The software tools used to calculate calibrated $S$-indices with spectra from ARCES are publicly available \replaced{\footnote{\url{https://github.com/bmorris3/arces_hk}}}{\footnote{\url{https://doi.org/10.5281/zenodo.886629}}}.

\subsection{The $S$-index of HAT-P-11} \label{sec:s_h11}

\begin{figure}
\centering
\includegraphics[scale=0.55]{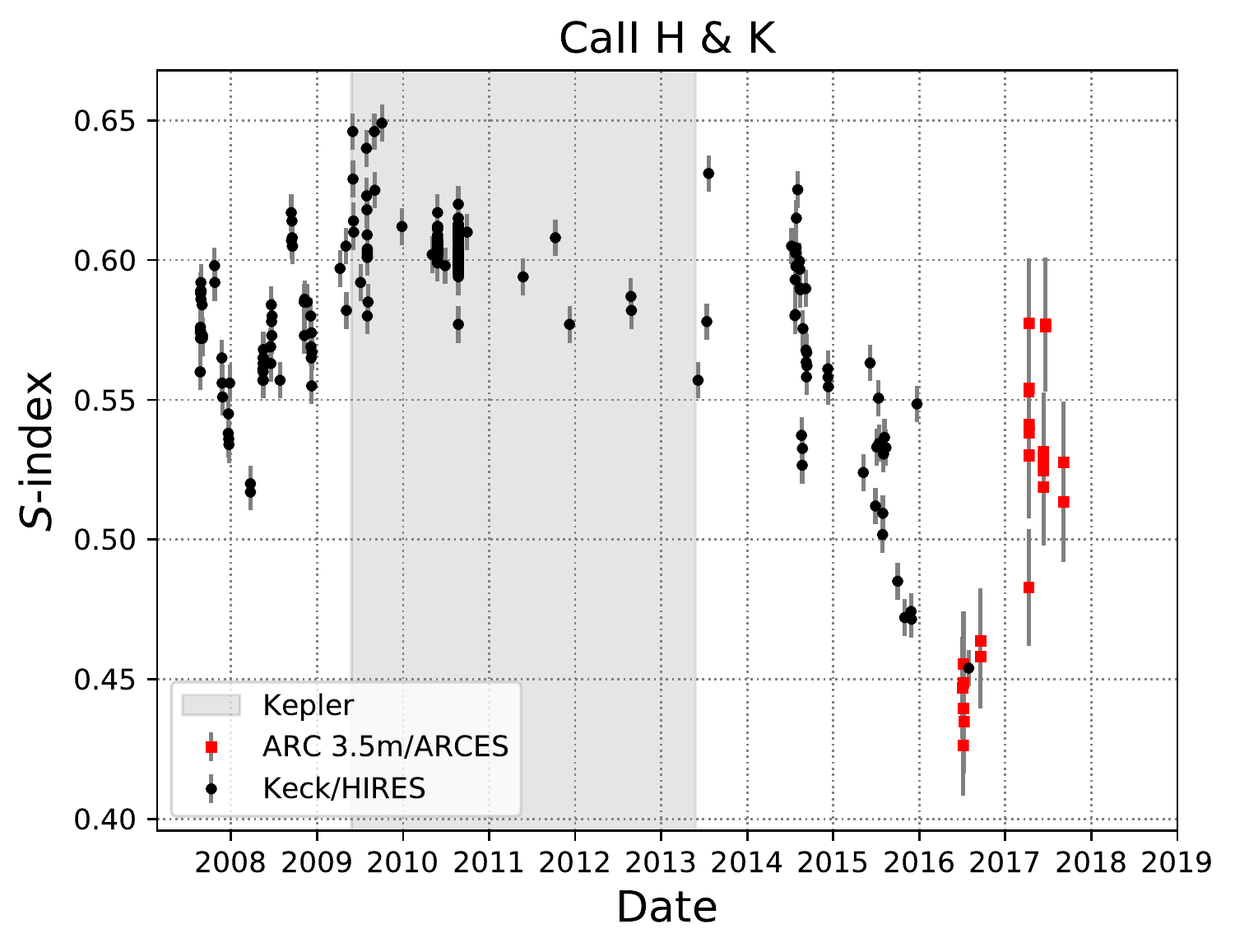}
\includegraphics[scale=0.55]{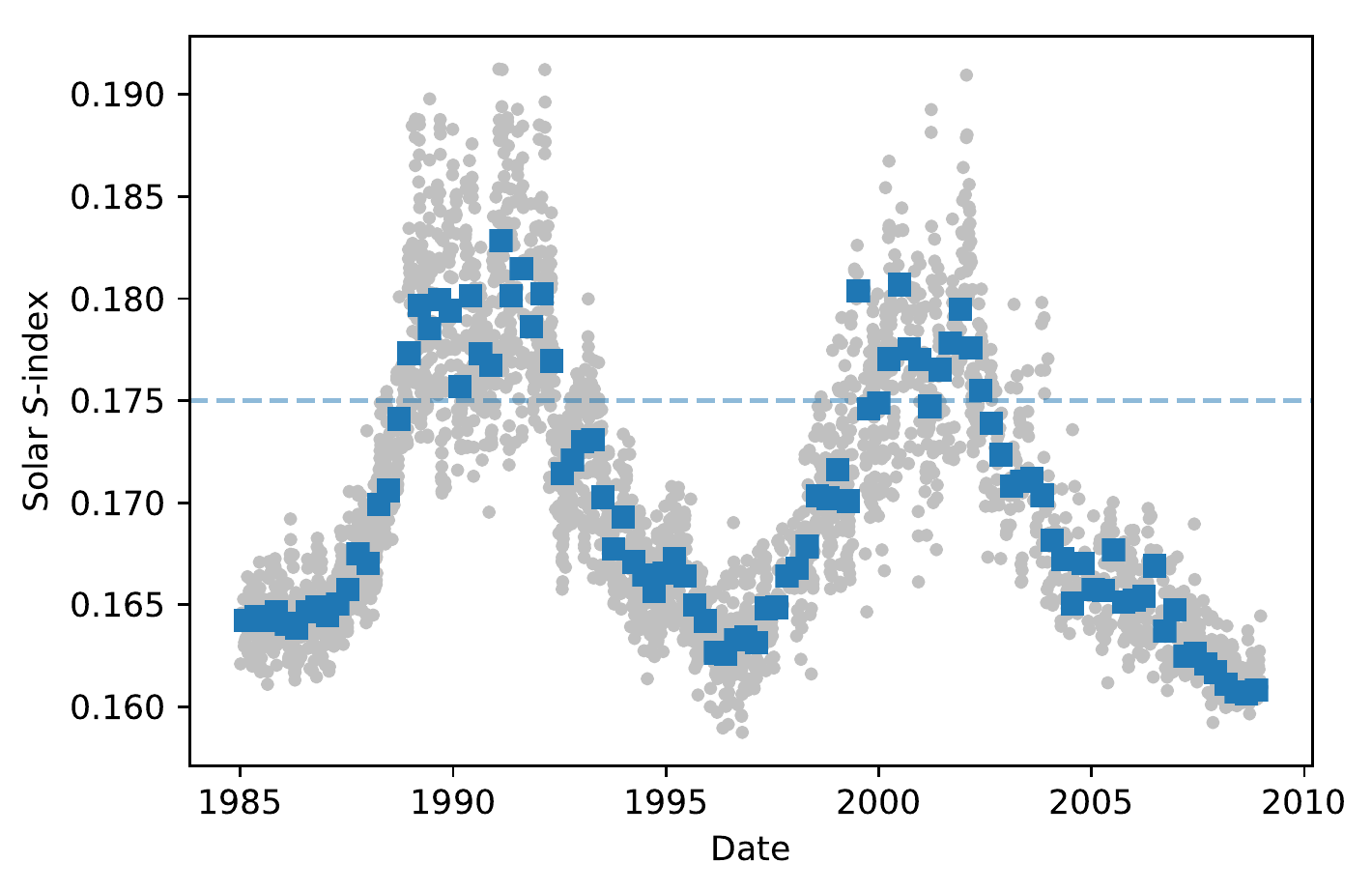}
\caption{\textit{Upper:} $S$-index of HAT-P-11 over time with archival Keck/HIRES spectra and calibrated APO spectra from this work. We see evidence for an activity cycle with period $\sim$10 years or more. \textit{Lower:} the $S$-index of the Sun, from the National Solar Observatory (NSO) Integrated Sunlight Spectrometer (ISS) on the Synoptic Optical Long-term Investigations of the Sun (SOLIS) telescope. We convert the ISS CaII $K$ line fluxes to $S$-indices with Eqns.~5 and 12 of \citet{egeland2017}. \added{If we have observed nearly one complete activity cycle of HAT-P-11, it appears that HAT-P-11 spends a longer fraction of the cycle near activity maximum than the Sun does.}}
\label{fig:activity_cycle}
\end{figure}

With the transformation computed in the previous section, we can compare the APO $S$-index measurements with those from the archival Keck/HIRES spectra. HAT-P-11 was the target of several Keck/HIRES programs since the planet's discovery, with the aims of measuring the Rossiter-McLaughlin effect and searching for a third body\footnote{We use archival spectra from program PIs: Bakos, D.~Bayliss, Beichman, Borucki, P.~Butler, D.~Fischer, Ford, Fortney, J.~Fuller, Gaidos, Hillenbrand, Howard, J~Johnson, Knutson, Mandushev, Marcy, Rogers, Sanchis-Ojeda, S.~Vogt}. We measure $S$-indices from these observations with the HIRES activity pipeline described in \citet{Isaacson2010}. A complete description of the HIRES pipeline developed for these measurements is beyond the scope of this paper; we refer the reader to \citet{Isaacson2010} for the full description. The result is an unevenly-sampled series of 239 $S$ measurements spanning 2007-2016, with typical $S/N \sim 20$. 

Figure~\ref{fig:activity_cycle} shows the $S$-index of HAT-P-11 between 2008-2017, including both the APO and Keck spectra. \added{The APO $S$-indices are enumerated in Table~\ref{tab:sind}}.The black circles are measurements from Keck/HIRES, and the red circles are measurements from APO. The chromospheric activity of HAT-P-11 appears to ramp up in the years preceding the \kepler\ mission, stabilize near a maximum throughout the \kepler\ years from 2009-2014, and then decline rapidly from 2014-2016 before rising at a similar rate through the present. 

Activity indices like $S$ vary on two timescales -- one driven by the rotation period of the star (29.2 d) as active regions rotate into and out of view, and the longer timescale throughout the activity cycle. It appears from this limited time series that HAT-P-11 may have entered activity maximum near the beginning of the \kepler\ observations. It is interesting to note that \citet{Morris2017} found that the starspots of HAT-P-11 are distributed in active latitudes centered on $\bar{\ell} = 16 \pm 1^\circ$, similar to the distribution of sunspots near solar activity maximum. The $S$-index appears to corroborate that if HAT-P-11's activity cycle is similar to the Sun's, it was at active maximum during the \kepler\ years.

Further spectroscopic monitoring will be required to determine the period of the activity cycle of HAT-P-11. During the summer of 2017, we measure $\left< S\right>_{2017} = 0.54 \pm 0.04$, still less than the earliest available measurements from the summer of 2007 $\left< S\right>_{2007} = 0.569 \pm 0.006$ -- it appears that one complete activity cycle has not yet elapsed. This gives a lower limit on the activity cycle period of $\sim 10$ years. The solar activity cycle is not perfectly periodic; the mean cycle duration and standard deviation are $10.9 \pm 1.2$ years \citep{Hathaway2002}. This lower limit on the activity cycle of HAT-P-11 observed thus far is longer than some solar activity cycles, but not longer than the mean. In principle one could attempt to measure the period of HAT-P-11's cycle by invoking empirical models of $S$ inspired by the solar activity cycle. However, as we will discuss below, HAT-P-11's $S$-index distribution with time departs from the typical $S$ distribution of the Sun -- HAT-P-11 spends more time near active maximum than minimum. Therefore we can only constrain a lower limit on the period with the observations thus far, and refrain from attempting to fit for the cycle period. 

The activity minimum of HAT-P-11 was short compared to its maximum. We note the marked difference between HAT-P-11 and the Sun in this respect in Figure~\ref{fig:activity_cycle}. Since we have not yet measured a full activity cycle of HAT-P-11 and the time sampling is very uneven, it is difficult to make concrete statements about the true duration of its maximum or minimum. One quantity that might be robust against these biases is the fraction of the activity cycle spent above or below the mean of $S_{min}$ and $S_{max}$, assuming the minimum and maximum observed $S$ are close to the true minimum and maximum of $S$ throughout the cycle. The Sun spends 25\% of the cycle above $0.5 (S_{min} + S_{max}) = 0.174$. The same quantity for HAT-P-11 is $0.5 (S_{min} + S_{max}) = 0.53$, and its $S$ was above that value from around or before 2008 until mid-2015, and stayed below that value until mid-2017. It appears that HAT-P-11 spent 63\% of its cycle above the mean of $S_{min}$ and $S_{max}$, assuming an $\sim 11$ year cycle. 

\added{If HAT-P-11 has an $\sim 11$ year activity cycle, it falls in a sparse region of rotation period-cycle period parameter space. \citet{Bohm-Vitense2007} showed that stars typically fall into one of two categories: the young, active $A$-sequence, and old, inactive $I$-sequence (see \citealt{Bohm-Vitense2007} Fig.~1). $A$-sequence stars typically have ratios of cycle periods to rotation periods $P_{cyc}/P_{rot} \sim 360$, and $I$-sequence stars have $P_{cyc}/P_{rot} \sim 80$.  The Sun is a notable outlier residing in between the sequences with $P_{cyc}/P_{rot} \sim 160$. If HAT-P-11's activity cycle is roughly $\sim 11$ years, it is interesting to note that it has $P_{cyc}/P_{rot} \sim 139$, and it falls between the $A$ and $I$ sequences, like the Sun.}

\begin{figure}
\begin{center}
\includegraphics[scale=0.55]{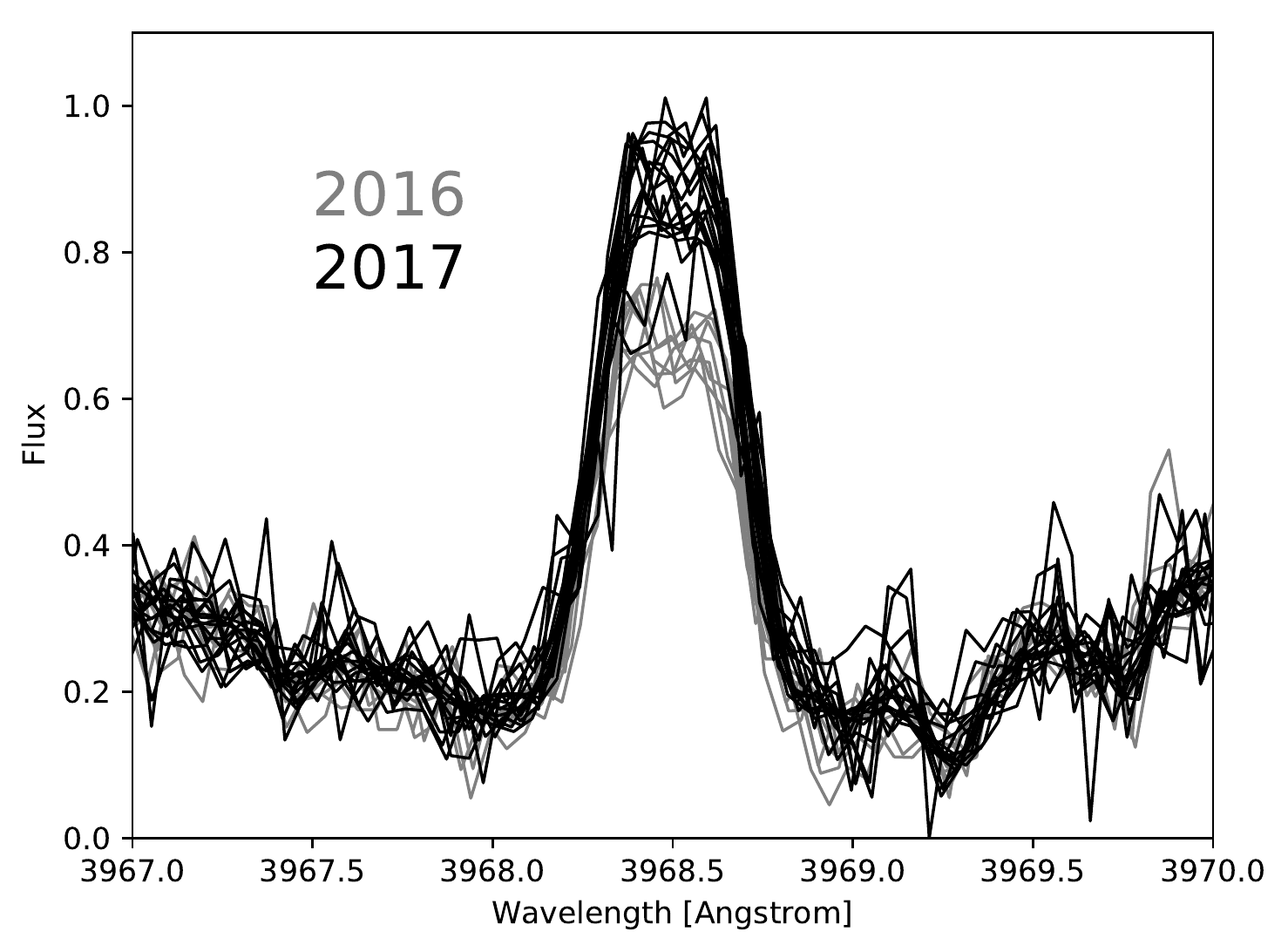}
\end{center}
\caption{The CaII H feature of HAT-P-11 shows clear evolution towards more activity from 2016 to 2017 in the APO spectra. It appears that the activity may be approaching its maximal level as of mid-2017.}
\label{fig:h}
\end{figure}

The APO spectra in Figure~\ref{fig:h} demonstrate that the activity has increased significantly in just the last two years. If we assume the cycle has a period of $\sim$11 years, we expect the next maximum to begin near 2019-2020. We predict that TESS photometry of HAT-P-11 will show spot occultations similar to those observed during the \kepler\ mission.

\section{\kepler\ Full-frame image photometry} \label{sec:ffi}

\citet{Montet2017} searched for long term variability of \kepler\ targets in photometry from the full-frame images (FFIs). These were special exposures where the entire \kepler\ detector was read-out, whereas the short and long cadence photometry from \kepler\ was only telemetered for a small subset of the pixels. There were eight FFIs in 2009 during commissioning, and then approximately one per month for the remainder of the mission.

\begin{figure}
\begin{center}
\includegraphics[scale=0.6]{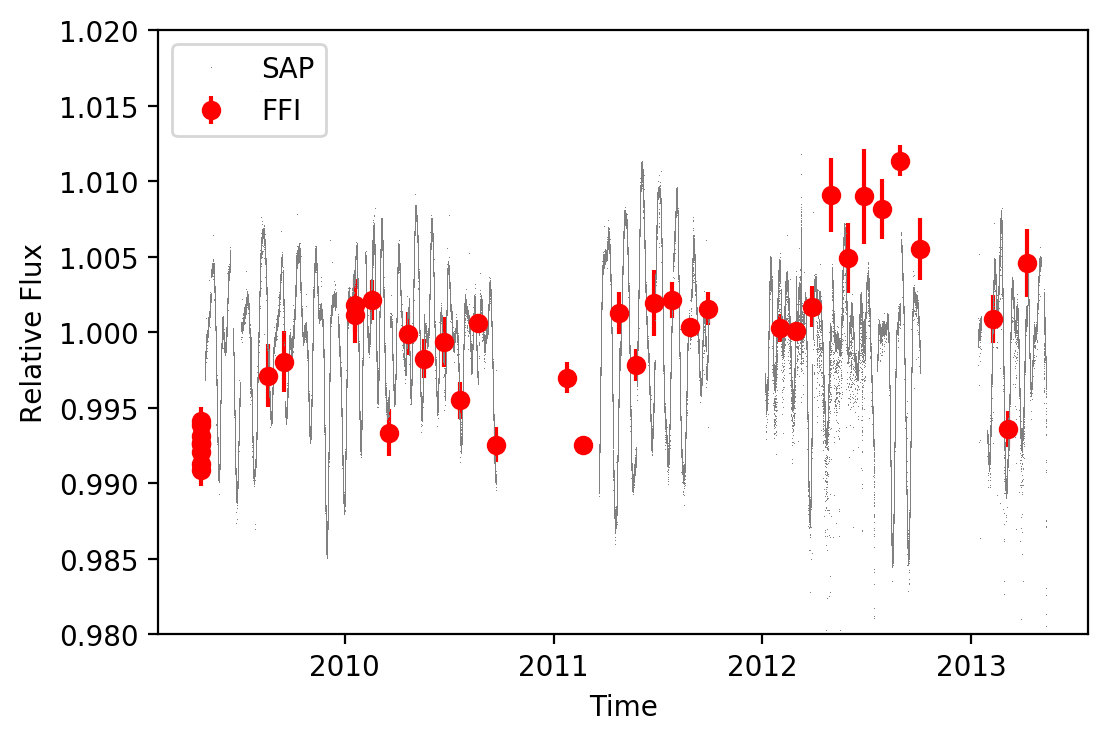}
\end{center}
\caption{Full-frame image photometry of HAT-P-11 (red circles) and \kepler\ SAP one minute cadence fluxes (gray). \citet{Montet2017} searched for evidence of stellar activity cycles in the full-frame image (FFI) photometry. The $S$-index was near maximum throughout the \kepler\ mission, so we expect the FFI photometry to show scatter consistent with the 2\% rotational modulation in the \kepler\ short-cadence light curve. The FFI photometry indeed show scatter consistent with the rotational variability, without a significant secular trend.}
\label{fig:ffi}
\end{figure}

We use HAT-P-11 as a test-case for the reliability of long-term trends in FFI photometry, see Figure~\ref{fig:ffi}. The FFI photometry shows that the flux of HAT-P-11 varied by $\sim 2\%$ from 2009-2013 without a significant secular trend. The scatter in the FFI photometry is consistent with the scatter in the short-cadence SAP fluxes due to rotational variability of the star. Note that one should only compare the global scatter of the FFI and SAP fluxes to one another, and one should not expect the SAP light curve to perfectly intersect with the SAP measurements, since the SAP fluxes shown in Figure~\ref{fig:ffi} are normalized by the median flux of each quarter, and the FFI fluxes are normalized by the median flux over all FFIs. 

The $S$-index in Figure~\ref{fig:activity_cycle} indicates that HAT-P-11 had nearly uniform chromospheric emission throughout the years of the \kepler\ mission. Both the \kepler\ photometry and $S$-index are consistent with an active maximum period from 2009-2013.

The long term photometric variability of most Sun-like stars with rotation periods of 29 d is dominated by bright faculae \citep{Montet2017}. Future space-based photometry missions could therefore expect to measure a small dimming of the star during its short activity minimum.

\section{HAT-P-11 in context} \label{sec:cks}

\subsection{Comparison to stars of similar color} \label{sec:mittag}

Is the chromospheric activity of HAT-P-11 typical for K4 dwarfs? The photometric analysis by \citet{Morris2017} can't be reproduced for many other \kepler\ stars. HAT-P-11 is exceptionally bright ($V=9.59$), and the highly inclined orbit of its planet reveals the latitude distribution of spots during transit events. Since very few similar targets are available for comparison in the \kepler\ sample, we seek to compare the spectrum of HAT-P-11 to spectra of other stars.

The $S$-index varies with stellar effective temperature, so the large $S$ of HAT-P-11 compared to the Sun does not imply that HAT-P-11 has more chromospheric activity than the Sun. The \rprime\ index is often used instead to compare activity levels across the main sequence, by normalizing the flux in the CaII H \& K emission features by the basal flux from the photosphere \citep{Noyes1984}. For stars of any effective temperature, \rprime\ increases with chromospheric activity.

We compute the \rprime\ index for a large sample of main sequence stars, following the procedure of \citet{Mittag2013}. We gathered published $S$-indices from \citet{Duncan1991}, \citet{Wright2004} and \citet{Isaacson2010}. As in \citet{Mittag2013}, we select only main sequence stars with color and absolute magnitude cuts using Hipparcos parallaxes \citep{Perryman1997}, yielding a sample of 4677 main sequence stars with measured $S$-indices. We then solve for \rprime\ for each star. We also compute \rprime\ for HAT-P-11 using its mean $\left < S \right> = 0.58 \pm 0.04$.

\begin{figure}
\begin{center}
\includegraphics[scale=0.8]{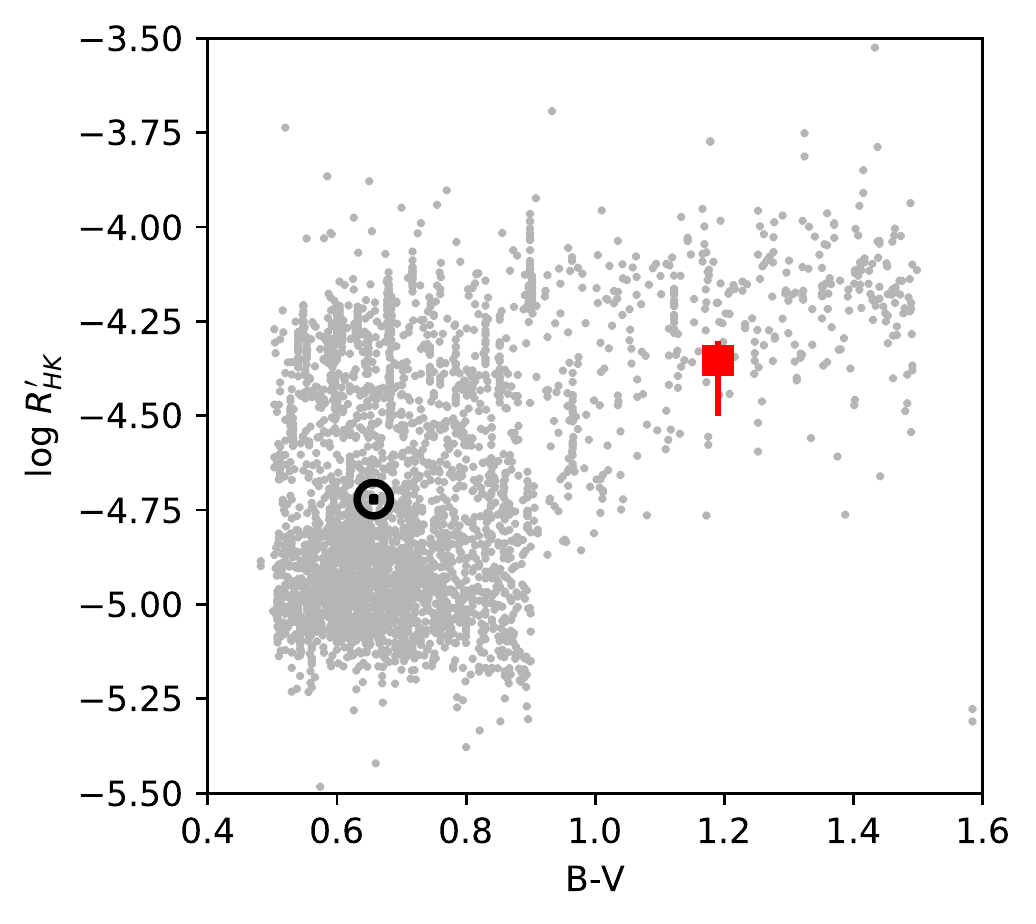}
\end{center}
\caption{The \rprime\ index of HAT-P-11 (red square) among other main sequence stars (circles). The Sun is represented by the black ``$\odot$'' symbol, and resides on the the inactive branch of bluer stars. The distribution of \rprime\ at a given $B-V$ color is dominated by variation with age. Dwarfs with $B-V > 1$ typically have more active chromospheres than dwarfs with $B-V <1$. The errorbars on the square representing HAT-P-11 denote the minimum and maximum \rprime\ given the complete range of observed $S$ values throughout the activity cycle.}
\label{fig:rprime_ms}
\end{figure}

The \rprime\ index for each star in the literature is shown in Figure~\ref{fig:rprime_ms}. The characteristic rise in chromospheric activity for later type stars is visible in the right half of the figure. HAT-P-11 is the red square with $B-V = 1.19$ and $\log R^\prime_{HK} = -4.35$ (evaluated for $\left<S\right>$). The Sun is represented with the $\odot$ symbol and resides in the inactive sequence, blueward of HAT-P-11. 

HAT-P-11 has typical chromospheric activity among field stars of similar color. It lies near the lower envelope of \rprime\ at $B-V\sim1.2$, which one might expect given its rotation period $P_{rot} = 29.2$ d since \rprime\ declines with increasing age and rotation period \citep{Noyes1984}. HAT-P-11 also hosts a close-in planet which one might speculate could affect its activity, and in general we don't know the planet occurrence frequency in this sample of stars. As a result, it is perhaps more interesting to compare the chromospheric activity of HAT-P-11 to planet-hosting stars of similar color \textit{and} age.

\subsection{Comparison to planet hosts of similar color and age}

\begin{figure}
\begin{center}
\includegraphics[scale=0.52]{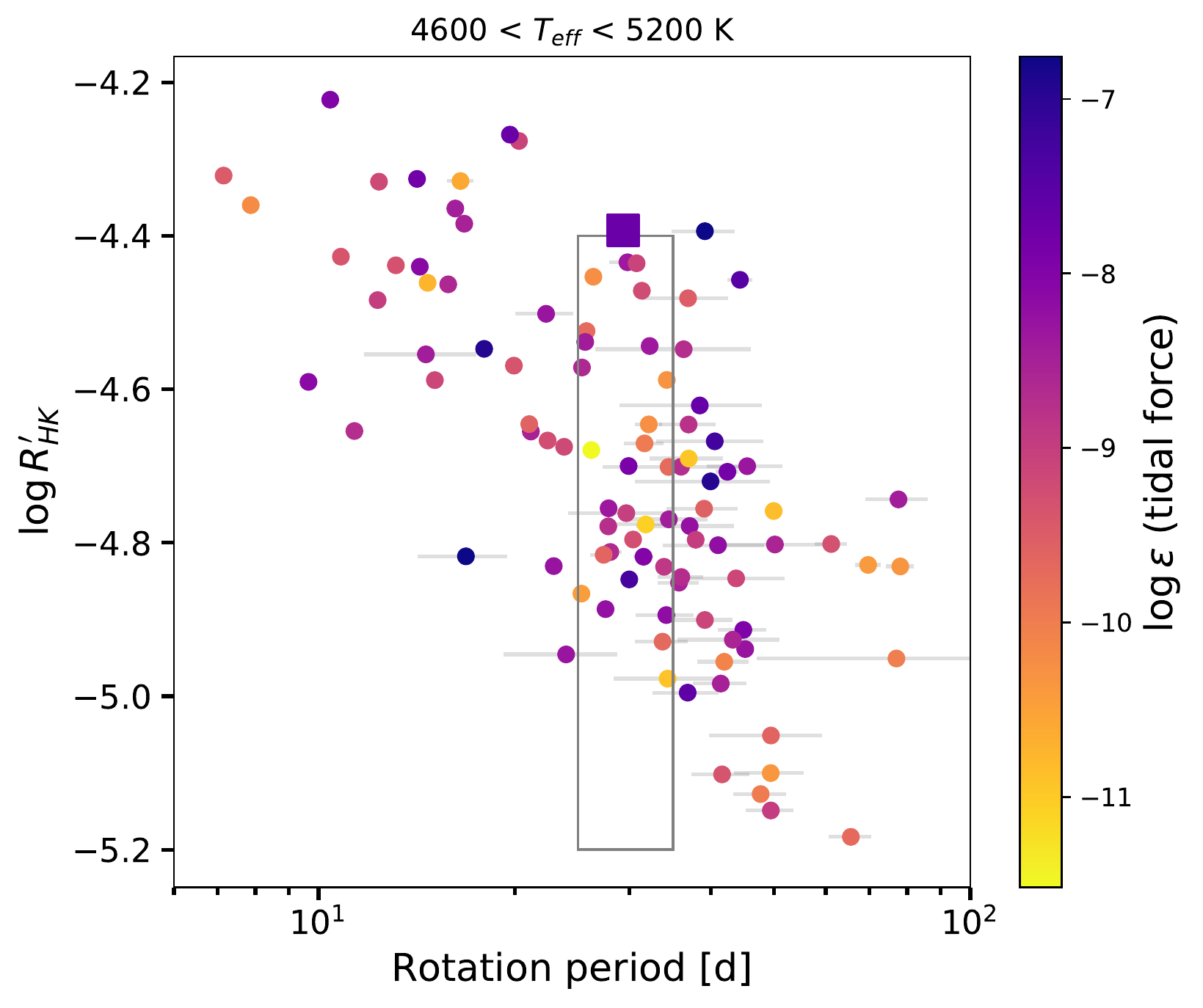}
\end{center}
\caption{The \rprime\ index of HAT-P-11 (big square) among other main sequence planet-hosting stars (circles). HAT-P-11 is among the most active planet-hosting K stars with rotation periods near 30 d -- its \rprime\ is similar to the median \rprime\ for stars with rotation periods near $\sim 15$ days.}
\label{fig:cks_activity}
\end{figure}

Is the chromospheric activity of HAT-P-11 typical for planet-hosting K stars with long rotation periods? We build a sample of stars with similar effective temperatures to HAT-P-11 from the California-Kepler Survey (CKS), in which \citet{Petigura2017} published Keck/HIRES spectra and \citet{Johnson2017} published precise stellar parameters for 1305 planet-hosting \kepler\ stars.  We gather the CKS spectra of 107 planet-hosting stars which meet the following criteria: they have (1) effective temperatures in the range $4600 < T_{eff} < 5200$ K \citet{Petigura2017}, bracketing HAT-P-11 at $T_{eff} = 4780 \pm 50$ K \citep{bakos2010}; (2) innermost planets with orbital periods $<20$ days; and (3) $S/N > 50$, where we approximate $S/N$ by dividing the median flux in the blue spectrum by the median uncertainty in the public CKS spectra.

We replicate the procedure in Section~\ref{sec:s_apo} to calculate $S$-indices for the CKS stars using the HIRES correction coefficients from \citet{Isaacson2010}, and use the relations of \citet{Mittag2013} to compute \rprime\ indices for each star as in Section~\ref{sec:mittag}. We also gather rotation periods for each star from \citet{Mazeh2015}, and we compute the Rossby number for each star with convective turnover times computed from the relations of \citet{Wright2011}.

We are particularly interested in whether or not HAT-P-11 experiences unusual planet-induced tides, which could provide a coupling between the planet's orbit and stellar activity. To quantify the relative tidal force on each star at the seat of stellar activity, the base of the convective zone, we use the simple dimensionless parameter $\epsilon$, 
\begin{equation}
\epsilon = \frac{M_p}{M_\star} \left( \frac{R_c}{a}\right)^3
\end{equation}
where $M_p$ and $M_\star$ are the masses of the planet and star, $R_c$ is the radius of the convective zone, and $a$ is planet's orbital semi-major axis \citep{Ogilvie2014}. The above equation is the ratio of the tidal force of the planet at the base of the convective zone $F_{tide} \propto M_p R_c / a^3$ to the local force of gravity within the star $F_g \propto M_\star / R_c^2$. This simple ratio does not take into account other likely important factors like the stellar obliquity. However, in the vast majority of cases the stellar obliquity is not known, so we continue to investigate tides with the imperfect index $\epsilon$, and discuss the implications of this caveat below.

We evaluate the tidal $\epsilon$ for the innermost planet in each system. We adopt the stellar radii and semi-major axes from the isochrone fits of \citet{Johnson2017}. We estimate the radius of the convective zone for each star by interpolating between the convective zone radii of model stars by \citet{vanSaders2012} of solar metallicity with $4600<T_{eff}<5200$ K. Finally, we adopt the mass measurement for HAT-P-11 b, $M_p=0.081 M_J$ from \citet{bakos2010}, and estimate the most probable mass for all other planets using the \texttt{forecaster} tool by \citet{Chen2017}, given the measured planetary radii from \citet{Johnson2017}.

We recover the canonical result that \rprime\ generally decreases with rotation rate (or age) in Figure~\ref{fig:cks_activity}. It appears that HAT-P-11 (large square) has more chromospheric activity than other p./,lanet-hosting stars of similar rotation periods and effective temperatures. The gray box in Figure~\ref{fig:cks_activity} circumscribes 31 stars with rotation periods $25 < P_{rot} < 35$ d. The median activity level within the box is only 40\% of HAT-P-11's, $\log R^\prime_{HK} = -4.75$.

\subsection{The possible role of tides}

\begin{figure}
\begin{center}
\includegraphics[scale=0.7]{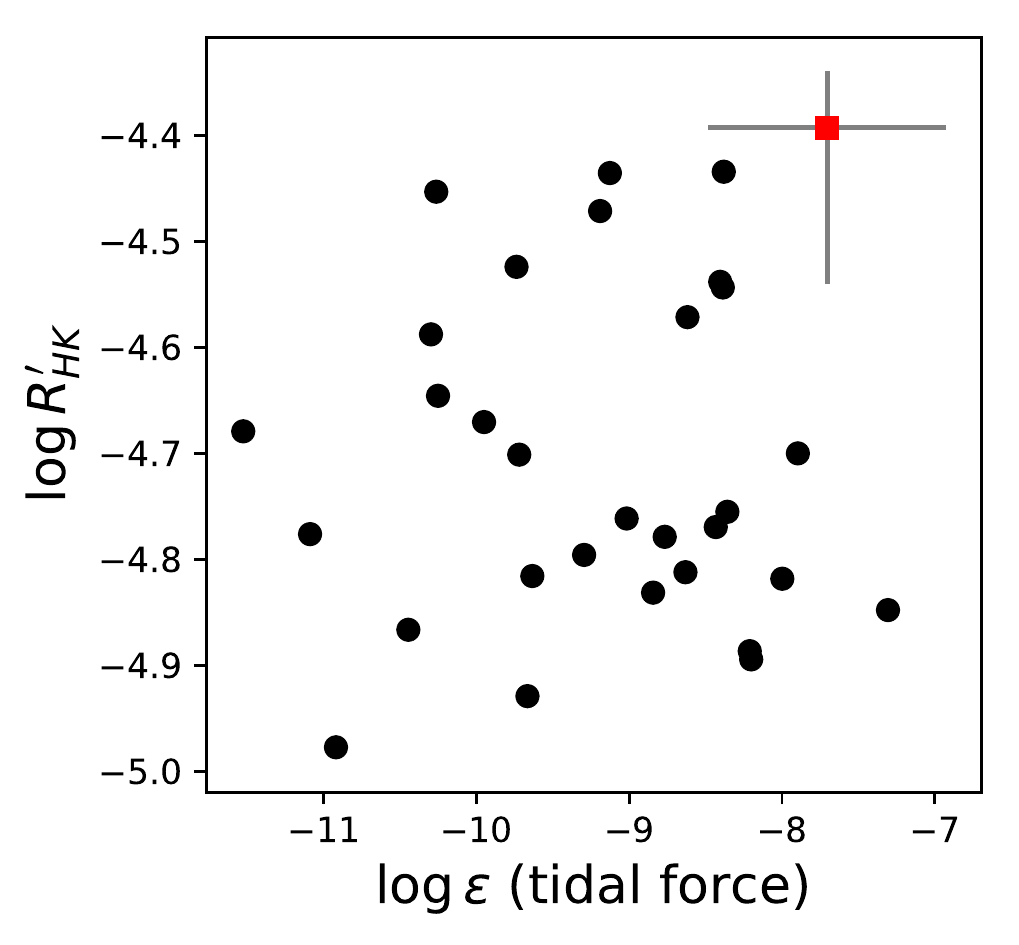}
\caption{The \rprime\ index for stars within the gray box in Figure~\ref{fig:cks_activity}, with rotation periods $25 < P_{rot} < 35$ d and effective temperatures $4600 < T_{eff} < 5200$ K. HAT-P-11 (red square) has the most activity in this bin, and the second-greatest tidal force on the base of its convective zone. \added{The uncertainty in \rprime\ on HAT-P-11 represent its variation throughout its activity cycle. The uncertainty in $\epsilon$ on HAT-P-11 represents the characteristic uncertainty for all systems, which is dominated by uncertainty in the planet mass-radius relation of \citet{Chen2017}.}}
\label{fig:rhk_eps}
\end{center}
\end{figure}

Could the relatively high level of chromospheric activity on HAT-P-11 (for stars with its age and color) be related to the tides raised by the close-in planet? We show the $\epsilon$-\rprime\ distribution of stars with $25 < P_{rot} < 35$ d in Figure~\ref{fig:rhk_eps}. We note that in this range of rotation periods, HAT-P-11 (red square) has the most chromospheric activity and the second-greatest $\epsilon$. However, the lack of  correlation between \rprime\ and $\epsilon$ suggests: (1) the tidal force isn't the only important factor in setting the level of chromospheric activity; and (2) the simple parameterization of $\epsilon$ does not capture all of the relevant physics. 

Significant tides are raised by HAT-P-11 b, even when compared to solar system objects. We can compare it's $\log_{10}\epsilon = -7.7$ with more familiar systems. $\epsilon$ at the base of the solar convective zone due to tides raised by Mercury gives $\log_{10}\epsilon = -13$, much less than any of the systems in Figure~\ref{fig:rhk_eps}. The Earth-Moon system experiences more similar tides with $\log_{10}\epsilon = -7.3$, and the dimensionless tidal force on HAT-P-11 is significantly less than the Jupiter-Io tides, $\log_{10}\epsilon = -6.6$. The tides raised by HAT-P-11 b are significant, and may have non-negligible effects on the interior dynamics of the star.

\subsection{The possible role of stellar obliquity}

Why is there no correlation between tidal force and chromospheric activity in this sample? We investigate the system with the greatest $\epsilon$ in more detail. That system is KOI-1050 with $\log_{10}\epsilon=-7.2$. Its star has mass $M_\star = 0.83 M_\odot$, effective temperature $T_{eff}= 5068$ K and rotation period $P_{rot} = 29.9$ d. It is orbited by planet with radius $R_p = 1.85 R_\oplus$, orbital period $1.27$ d and forecasted mass $M_p = 4.4 M_\oplus$. \rprime\ of KOI-1050 is less than half of HAT-P-11's despite its larger $\epsilon$. 

The dimensionless tidal force $\epsilon$ does not account for the transfer of energy due to obliquity tides -- the result of additional torques in star-planet systems with planetary orbits that are highly misaligned with respect to the stellar spin. The amplitude of energy transfer due to obliquity tides is proportional to $\sin^2 \psi$, where $\psi$ is the angle between the orbital angular moment vector and the stellar spin axis \citep{Wisdom2004}. The high obliquity of HAT-P-11 $\psi = 106^\circ {}^{+15}_{-11}$ suggests that obliquity tides on HAT-P-11 will contribute additional dissipation of orbital energy, which is not reflected by $\epsilon$. Lacking measurements of the obliquity of the other systems in Figure~\ref{fig:rhk_eps} such as KOI-1050, the observations presented here are insufficient to comment on whether or not the high obliquity of HAT-P-11 is responsible for its greater \rprime\ activity index compared to KOI-1050, for example. Future measurements of spin-orbit misalignment of close-in planets with active host stars may allow us to link stellar activity to the tides raised by planets.

\added{It is also possible that a weak correlation between $\epsilon$ and \rprime\ is obscured by the large uncertainties in the forecasted planet masses. The parameter with the largest uncertainty in the calculation for $\epsilon$ is the planet mass. The forecasted mass estimates from \citet{Chen2017} have large uncertainties which reflect both measurement uncertainty and intrinsic spread in the observed planetary mass-radius relation. For the planets in the sample considered here, the median uncertainty in the forecasted planet mass is 78\%, which contributes to uncertainty on the order of 1 dex in epsilon. Thus the typical uncertainty in our estimates of $\epsilon$ is significant compared to the range observed. Follow-up radial velocity or transit timing variation measurements are needed to better constrain planet masses, and thus their tidal influence on their host stars.}

\added{Alternative activity indicators have been used to search for evidence that tidal interactions stoke stellar activity \citep[see reviews by][]{wright2015, poppenhaeger2017}. For example, \citet{Poppenhaeger2014} and \citet{miller2015} found enhanced stellar coronal X-ray emission from host stars that are expected to have strong tidal interactions with their hot Jupiters. \citet{saar2001} measured the chromospheric emission in the CaII infrared-triplet over time for stars with hot Jupiters, and found that the emission did not fluctuate on the orbital period of the planet.

In addition to tidal interactions, close-in planets can also interact magnetically with their host stars. \citet{Cohen2010, Lanza2010} showed that star-planet magnetic interactions can affect the angular momentum evolution of host stars. These studies examined planets with orbit normals that were aligned or anti-aligned with the stellar spin. The magnetic interactions between the star and planet in the HAT-P-11 system -- with its planet in a polar orbit -- are worthy of further study, and beyond the scope of this paper.}

\section{Conclusion}

We present $S$-indices for HAT-P-11 from 2008-2017, which show evidence of an activity cycle of duration $\gtrsim 10$ years. We detail the calibration procedure for measuring $S$-indices with the echelle spectrograph on the ARC 3.5 m Telescope at APO. If the activity cycle is $\sim 11$ years, we expect the star to remain highly active from the present until $\sim 2023$, thus TESS will be able to observe spot occultations throughout the active maximum.

If we interpret the local minimum in $S$ near mid-2016 to be the true activity minimum, it seems that HAT-P-11 spends a greater fraction of time at maximum ($\sim 63\%$)  than the Sun does ($\sim 25\%$). The brightness of HAT-P-11 throughout the maximum lasting from 2009-2014 is consistent with under-sampled rotational variability, and does not show a significant secular trend. 

Among all K dwarfs, the strength of HAT-P-11's chromospheric activity measured by \rprime\ is unremarkable. However, among planet-hosting dwarfs with similar rotation periods and effective temperatures, the chromosphere of HAT-P-11 is exceptionally active. We speculate that tides raised by the planet on the star may play a role in the atypical activity.

\acknowledgments

We gratefully acknowledge support from NSF grant AST-1312453. We thank Lauren Weiss, John Lurie, and Daniel Foreman-Mackey for helpful discussions. 

JRAD is supported by an NSF Astronomy and Astrophysics Postdoctoral Fellowship under award AST-1501418. 

Work by B.T.M. was performed under contract with the California Institute of Technology (Caltech)/Jet Propulsion Laboratory (JPL) funded by NASA through the Sagan Fellowship Program executed by the NASA Exoplanet Science Institute. C.M.S. acknowledges funding from the Kenilworth Fund of the New York Community Trust.

Based on observations obtained with the Apache Point Observatory 3.5-meter telescope, which is owned and operated by the Astrophysical Research Consortium. IRAF is distributed by the National Optical Astronomy Observatory, which is operated by the Association of Universities for Research in Astronomy, Inc., under cooperative agreement with the National Science Foundation

Some of the data presented herein were obtained at the W. M. Keck Observatory, which is operated as a scientific partnership among the California Institute of Technology, the University of California and the National Aeronautics and Space Administration. The Observatory was made possible by the generous financial support of the W. M. Keck Foundation. The authors wish to recognize and acknowledge the very significant cultural role and reverence that the summit of Maunakea has always had within the indigenous Hawaiian community.  We are most fortunate to have the opportunity to conduct observations from this mountain.

\facility{KeckI:HIRES, APO/ARC, Kepler}

\software{\texttt{ipython} \citep{ipython}, \texttt{numpy} \citep{VanDerWalt2011}, \texttt{scipy} \citep{scipy},  \texttt{matplotlib} \citep{matplotlib}, \texttt{astropy} \citep{Astropy2013}, \texttt{forecaster} \citep{Chen2017}}


\begin{thebibliography}{}

\bibitem[\protect\astroncite{{Astropy Collaboration}
  et~al.}{2013}]{Astropy2013}
{Astropy Collaboration}, T.~P. {Robitaille}, E.~J. {Tollerud}, P.~{Greenfield},
  M.~{Droettboom}, E.~{Bray}, T.~{Aldcroft}, M.~{Davis}, A.~{Ginsburg}, A.~M.
  {Price-Whelan}, W.~E. {Kerzendorf}, A.~{Conley}, N.~{Crighton}, K.~{Barbary},
  D.~{Muna}, H.~{Ferguson}, F.~{Grollier}, M.~M. {Parikh}, P.~H. {Nair}, H.~M.
  {Unther}, C.~{Deil}, J.~{Woillez}, S.~{Conseil}, R.~{Kramer}, J.~E.~H.
  {Turner}, L.~{Singer}, R.~{Fox}, B.~A. {Weaver}, V.~{Zabalza}, Z.~I.
  {Edwards}, K.~{Azalee Bostroem}, D.~J. {Burke}, A.~R. {Casey}, S.~M.
  {Crawford}, N.~{Dencheva}, J.~{Ely}, T.~{Jenness}, K.~{Labrie}, P.~{Lian
  Lim}, F.~{Pierfederici}, A.~{Pontzen}, A.~{Ptak}, B.~{Refsdal},
  M.~{Servillat}, and O.~{Streicher}\leavevmode\nopagebreak\newline 2013.
\newblock {Astropy: A community Python package for astronomy}.
\newblock {\em \aap}, 558:A33.

\bibitem[\protect\astroncite{Jones et~al.}{01  }]{scipy}
Jones, E., T.~Oliphant, P.~Peterson, et~al.\leavevmode\nopagebreak\newline
  2001--.
\newblock {SciPy}: Open source scientific tools for {Python}.

\bibitem[\protect\astroncite{{Hunter}}{2007}]{matplotlib}
{Hunter}, J.~D.\leavevmode\nopagebreak\newline 2007.
\newblock {Matplotlib: A 2D Graphics Environment}.
\newblock {\em Computing in Science and Engineering}, 9:90--95.

\bibitem[\protect\astroncite{{Van Der Walt} et~al.}{2011}]{VanDerWalt2011}
{Van Der Walt}, S., S.~C. {Colbert}, and
  G.~{Varoquaux}\leavevmode\nopagebreak\newline 2011.
\newblock {The NumPy array: a structure for efficient numerical computation}.
\newblock {\em ArXiv e-prints}.

\bibitem[\protect\astroncite{Perez and Granger}{2007}]{ipython}
Perez, F. and B.~E. Granger\leavevmode\nopagebreak\newline 2007.
\newblock Ipython: A system for interactive scientific computing.
\newblock {\em Computing in Science and Engg.}, 9(3):21--29.

\bibitem[\protect\astroncite{{Bakos} et~al.}{2010}]{bakos2010}
{Bakos}, G.~{\'A}., G.~{Torres}, A.~{P{\'a}l}, J.~{Hartman}, G.~{Kov{\'a}cs},
  R.~W. {Noyes}, D.~W. {Latham}, D.~D. {Sasselov}, B.~{Sip{\H o}cz}, G.~A.
  {Esquerdo}, D.~A. {Fischer}, J.~A. {Johnson}, G.~W. {Marcy}, R.~P. {Butler},
  H.~{Isaacson}, A.~{Howard}, S.~{Vogt}, G.~{Kov{\'a}cs}, J.~{Fernandez},
  A.~{Mo{\'o}r}, R.~P. {Stefanik}, J.~{L{\'a}z{\'a}r}, I.~{Papp}, and
  P.~{S{\'a}ri}\leavevmode\nopagebreak\newline 2010.
\newblock {HAT-P-11b: A Super-Neptune Planet Transiting a Bright K Star in the
  Kepler Field}.
\newblock {\em \apj}, 710:1724--1745.

\bibitem[\protect\astroncite{{Baliunas} et~al.}{1995}]{Baliunas1995}
{Baliunas}, S.~L., R.~A. {Donahue}, W.~H. {Soon}, J.~H. {Horne}, J.~{Frazer},
  L.~{Woodard-Eklund}, M.~{Bradford}, L.~M. {Rao}, O.~C. {Wilson}, Q.~{Zhang},
  W.~{Bennett}, J.~{Briggs}, S.~M. {Carroll}, D.~K. {Duncan}, D.~{Figueroa},
  H.~H. {Lanning}, T.~{Misch}, J.~{Mueller}, R.~W. {Noyes}, D.~{Poppe}, A.~C.
  {Porter}, C.~R. {Robinson}, J.~{Russell}, J.~C. {Shelton}, T.~{Soyumer},
  A.~H. {Vaughan}, and J.~H. {Whitney}\leavevmode\nopagebreak\newline 1995.
\newblock {Chromospheric variations in main-sequence stars}.
\newblock {\em \apj}, 438:269--287.

\bibitem[B{\"o}hm-Vitense(2007)]{Bohm-Vitense2007} B{\"o}hm-Vitense, E.\ 2007, \apj, 657, 486 

\bibitem[\protect\astroncite{{Brown} et~al.}{2011}]{Brown2011}
{Brown}, T.~M., D.~W. {Latham}, M.~E. {Everett}, and G.~A.
  {Esquerdo}\leavevmode\nopagebreak\newline 2011.
\newblock {Kepler Input Catalog: Photometric Calibration and Stellar
  Classification}.
\newblock {\em \aj}, 142:112.

\bibitem[\protect\astroncite{{Chen} and {Kipping}}{2017}]{Chen2017}
{Chen}, J. and D.~{Kipping}\leavevmode\nopagebreak\newline 2017.
\newblock {Probabilistic Forecasting of the Masses and Radii of Other Worlds}.
\newblock {\em \apj}, 834:17.

\bibitem[\protect\astroncite{{Duncan} et~al.}{1991}]{Duncan1991}
{Duncan}, D.~K., A.~H. {Vaughan}, O.~C. {Wilson}, G.~W. {Preston}, J.~{Frazer},
  H.~{Lanning}, A.~{Misch}, J.~{Mueller}, D.~{Soyumer}, L.~{Woodard}, S.~L.
  {Baliunas}, R.~W. {Noyes}, L.~W. {Hartmann}, A.~{Porter}, C.~{Zwaan},
  F.~{Middelkoop}, R.~G.~M. {Rutten}, and
  D.~{Mihalas}\leavevmode\nopagebreak\newline 1991.
\newblock {CA II H and K measurements made at Mount Wilson Observatory,
  1966-1983}.
\newblock {\em \apjs}, 76:383--430.

\bibitem[\protect\astroncite{{Egeland} et~al.}{2017}]{egeland2017}
{Egeland}, R., W.~{Soon}, S.~{Baliunas}, J.~C. {Hall}, A.~A. {Pevtsov}, and
  L.~{Bertello}\leavevmode\nopagebreak\newline 2017.
\newblock {The Mount Wilson Observatory S-index of the Sun}.
\newblock {\em \apj}, 835:25.

\bibitem[\protect\astroncite{{Foreman-Mackey}
  et~al.}{2013}]{Foreman-Mackey2013}
{Foreman-Mackey}, D., D.~W. {Hogg}, D.~{Lang}, and
  J.~{Goodman}\leavevmode\nopagebreak\newline 2013.
\newblock {emcee: The MCMC Hammer}.
\newblock {\em \pasp}, 125:306--312.

\bibitem[\protect\astroncite{{Goodman} and {Weare}}{2010}]{Goodman2010}
{Goodman}, J. and J.~{Weare}\leavevmode\nopagebreak\newline 2010.
\newblock {Ensemble samplers with affine invariance}.
\newblock {\em Communications in Applied Mathematics and Computational
  Science}, 5:65--80.

\bibitem[\protect\astroncite{{Hall}}{2008}]{Hall2008}
{Hall}, J.~C.\leavevmode\nopagebreak\newline 2008.
\newblock {Stellar Chromospheric Activity}.
\newblock {\em Living Reviews in Solar Physics}, 5:2.

\bibitem[\protect\astroncite{{Hathaway} et~al.}{2002}]{Hathaway2002}
{Hathaway}, D.~H., R.~M. {Wilson}, and E.~J.
  {Reichmann}\leavevmode\nopagebreak\newline 2002.
\newblock {Group Sunspot Numbers: Sunspot Cycle Characteristics}.
\newblock {\em \solphys}, 211:357--370.

\bibitem[\protect\astroncite{{Husser} et~al.}{2013}]{Husser2013}
{Husser}, T.-O., S.~{Wende-von Berg}, S.~{Dreizler}, D.~{Homeier},
  A.~{Reiners}, T.~{Barman}, and P.~H.
  {Hauschildt}\leavevmode\nopagebreak\newline 2013.
\newblock {A new extensive library of PHOENIX stellar atmospheres and synthetic
  spectra}.
\newblock {\em \aap}, 553:A6.

\bibitem[\protect\astroncite{{Isaacson} and {Fischer}}{2010}]{Isaacson2010}
{Isaacson}, H. and D.~{Fischer}\leavevmode\nopagebreak\newline 2010.
\newblock {Chromospheric Activity and Jitter Measurements for 2630 Stars on the
  California Planet Search}.
\newblock {\em \apj}, 725:875--885.

\bibitem[\protect\astroncite{{Johnson} et~al.}{2017}]{Johnson2017}
{Johnson}, J.~A., E.~A. {Petigura}, B.~J. {Fulton}, G.~W. {Marcy}, A.~W.
  {Howard}, H.~{Isaacson}, L.~{Hebb}, P.~A. {Cargile}, T.~D. {Morton}, L.~M.
  {Weiss}, J.~N. {Winn}, L.~A. {Rogers}, E.~{Sinukoff}, and L.~A.
  {Hirsch}\leavevmode\nopagebreak\newline 2017.
\newblock {The California-Kepler Survey. II. Precise Physical Properties of
  2025 Kepler Planets and Their Host Stars}.
\newblock {\em ArXiv e-prints}.

\bibitem[\protect\astroncite{{Mazeh} et~al.}{2015}]{Mazeh2015}
{Mazeh}, T., H.~B. {Perets}, A.~{McQuillan}, and E.~S.
  {Goldstein}\leavevmode\nopagebreak\newline 2015.
\newblock {Photometric Amplitude Distribution of Stellar Rotation of KOIs -
  Indication for Spin-Orbit Alignment of Cool Stars and High Obliquity for Hot
  Stars}.
\newblock {\em \apj}, 801:3.

\bibitem[\protect\astroncite{{Mittag} et~al.}{2013}]{Mittag2013}
{Mittag}, M., J.~H.~M.~M. {Schmitt}, and K.-P.
  {Schr{\"o}der}\leavevmode\nopagebreak\newline 2013.
\newblock {Ca II H+K fluxes from S-indices of large samples: a reliable and
  consistent conversion based on PHOENIX model atmospheres}.
\newblock {\em \aap}, 549:A117.

\bibitem[\protect\astroncite{{Montet} et~al.}{2017}]{Montet2017}
{Montet}, B.~T., G.~{Tovar}, and
  D.~{Foreman-Mackey}\leavevmode\nopagebreak\newline 2017.
\newblock {Long Term Photometric Variability in Kepler Full Frame Images:
  Magnetic Cycles of Sun-Like Stars}.
\newblock {\em ArXiv e-prints}.

\bibitem[\protect\astroncite{Morris et~al.}{2017}]{Morris2017}
Morris, B.~M., L.~Hebb, J.~R.~A. Davenport, G.~Rohn, and S.~L.
  Hawley\leavevmode\nopagebreak\newline 2017.
\newblock The starspots of hat-p-11: Evidence for a solar-like dynamo.
\newblock {\em The Astrophysical Journal}, 846(2):99.

\bibitem[\protect\astroncite{{Noyes} et~al.}{1984}]{Noyes1984}
{Noyes}, R.~W., L.~W. {Hartmann}, S.~L. {Baliunas}, D.~K. {Duncan}, and A.~H.
  {Vaughan}\leavevmode\nopagebreak\newline 1984.
\newblock {Rotation, convection, and magnetic activity in lower main-sequence
  stars}.
\newblock {\em \apj}, 279:763--777.

\bibitem[\protect\astroncite{{Ogilvie}}{2014}]{Ogilvie2014}
{Ogilvie}, G.~I.\leavevmode\nopagebreak\newline 2014.
\newblock {Tidal Dissipation in Stars and Giant Planets}.
\newblock {\em \araa}, 52:171--210.

\bibitem[\protect\astroncite{{Perryman} et~al.}{1997}]{Perryman1997}
{Perryman}, M.~A.~C., L.~{Lindegren}, J.~{Kovalevsky}, E.~{Hoeg}, U.~{Bastian},
  P.~L. {Bernacca}, M.~{Cr{\'e}z{\'e}}, F.~{Donati}, M.~{Grenon}, M.~{Grewing},
  F.~{van Leeuwen}, H.~{van der Marel}, F.~{Mignard}, C.~A. {Murray}, R.~S. {Le
  Poole}, H.~{Schrijver}, C.~{Turon}, F.~{Arenou}, M.~{Froeschl{\'e}}, and
  C.~S. {Petersen}\leavevmode\nopagebreak\newline 1997.
\newblock {The HIPPARCOS Catalogue}.
\newblock {\em \aap}, 323:L49--L52.

\bibitem[\protect\astroncite{{Petigura} et~al.}{2017}]{Petigura2017}
{Petigura}, E.~A., A.~W. {Howard}, G.~W. {Marcy}, J.~A. {Johnson},
  H.~{Isaacson}, P.~A. {Cargile}, L.~{Hebb}, B.~J. {Fulton}, L.~M. {Weiss},
  T.~D. {Morton}, J.~N. {Winn}, L.~A. {Rogers}, E.~{Sinukoff}, L.~A. {Hirsch},
  and I.~J.~M. {Crossfield}\leavevmode\nopagebreak\newline 2017.
\newblock {The California-Kepler Survey. I. High Resolution Spectroscopy of
  1305 Stars Hosting Kepler Transiting Planets}.
\newblock {\em ArXiv e-prints}.

\bibitem[\protect\astroncite{{Skumanich}}{1972}]{Skumanich1972}
{Skumanich}, A.\leavevmode\nopagebreak\newline 1972.
\newblock {Time Scales for CA II Emission Decay, Rotational Braking, and
  Lithium Depletion}.
\newblock {\em \apj}, 171:565.

\bibitem[\protect\astroncite{{van Saders} and
  {Pinsonneault}}{2012}]{vanSaders2012}
{van Saders}, J.~L. and M.~H. {Pinsonneault}\leavevmode\nopagebreak\newline
  2012.
\newblock {The Sensitivity of Convection Zone Depth to Stellar Abundances: An
  Absolute Stellar Abundance Scale from Asteroseismology}.
\newblock {\em \apj}, 746:16.

\bibitem[\protect\astroncite{{Wilson}}{1978}]{Wilson1978}
{Wilson}, O.~C.\leavevmode\nopagebreak\newline 1978.
\newblock {Chromospheric variations in main-sequence stars}.
\newblock {\em \apj}, 226:379--396.

\bibitem[\protect\astroncite{{Wisdom}}{2004}]{Wisdom2004}
{Wisdom}, J.\leavevmode\nopagebreak\newline 2004.
\newblock {Spin-Orbit Secondary Resonance Dynamics of Enceladus}.
\newblock {\em \aj}, 128:484--491.

\bibitem[\protect\astroncite{{Wright} et~al.}{2004}]{Wright2004}
{Wright}, J.~T., G.~W. {Marcy}, R.~P. {Butler}, and S.~S.
  {Vogt}\leavevmode\nopagebreak\newline 2004.
\newblock {Chromospheric Ca II Emission in Nearby F, G, K, and M Stars}.
\newblock {\em \apjs}, 152:261--295.

\bibitem[\protect\astroncite{{Wright} et~al.}{2011}]{Wright2011}
{Wright}, N.~J., J.~J. {Drake}, E.~E. {Mamajek}, and G.~W.
  {Henry}\leavevmode\nopagebreak\newline 2011.
\newblock {The Stellar-activity-Rotation Relationship and the Evolution of
  Stellar Dynamos}.
\newblock {\em \apj}, 743:48.

\bibitem[\protect\astroncite{{Miller} et~al.}{2015}]{miller2015}
{Miller}, B.~P., E.~{Gallo}, J.~T. {Wright}, and E.~G.
  {Pearson}\leavevmode\nopagebreak\newline 2015.
\newblock {A Comprehensive Statistical Assessment of Star-Planet Interaction}.
\newblock {\em \apj}, 799:163.

\bibitem[\protect\astroncite{{Poppenhaeger} and
  {Wolk}}{2014}]{Poppenhaeger2014}
{Poppenhaeger}, K. and S.~J. {Wolk}\leavevmode\nopagebreak\newline 2014.
\newblock {Indications for an influence of hot Jupiters on the rotation and
  activity of their host stars}.
\newblock {\em \aap}, 565:L1.

\bibitem[\protect\astroncite{{Saar} and {Cuntz}}{2001}]{saar2001}
{Saar}, S.~H. and M.~{Cuntz}\leavevmode\nopagebreak\newline 2001.
\newblock {A search for Ca II emission enhancement in stars resulting from
  nearby giant planets}.
\newblock {\em \mnras}, 325:55--59.

\bibitem[\protect\astroncite{Wright and Miller}{2015}]{wright2015}
Wright, J.~T. and B.~P. Miller\leavevmode\nopagebreak\newline 2015.
\newblock Magnetism and activity of planet hosting stars.
\newblock {\em Proceedings of the International Astronomical Union},
  11(S320):357–366.
  
\bibitem[\protect\astroncite{{Poppenhaeger}}{2017}]{poppenhaeger2017}
{Poppenhaeger}, K.\leavevmode\nopagebreak\newline 2017.
\newblock {Tidal effects on stellar activity}.
\newblock {\em ArXiv e-prints}.

\bibitem[\protect\astroncite{{Cohen} et~al.}{2010}]{Cohen2010}
{Cohen}, O., J.~J. {Drake}, V.~L. {Kashyap}, I.~V. {Sokolov}, and T.~I.
  {Gombosi}\leavevmode\nopagebreak\newline 2010.
\newblock {The Impact of Hot Jupiters on the Spin-down of their Host Stars}.
\newblock {\em \apjl}, 723:L64--L67.

\bibitem[\protect\astroncite{{Lanza}}{2010}]{Lanza2010}
{Lanza}, A.~F.\leavevmode\nopagebreak\newline 2010.
\newblock {Hot Jupiters and the evolution of stellar angular momentum}.
\newblock {\em \aap}, 512:A77.

\end{thebibliography}

\appendix

\begin{table}[h]
\begin{center}
\caption{Stars observed to calibrate the $S$-index. \label{tab:cals}}
\begin{tabular}{l l c c c}
Star & Sp.~Type & $S_{MWO}$ & $S_{APO}$ & $N$ \\
\hline
HD 210905 & K0III & $0.092 \pm 0.013$ & $0.004 \pm 0.000027$ & 1 \\
HD34411 & G1V & $0.145 \pm 0.022$ & $0.007 \pm 0.000069$ & 2 \\
HD68017 & G3V & $0.174 \pm 0.024$ & $0.008 \pm 0.00011$ & 2 \\
HD98230 & G2V & $0.266 \pm 0.031$ & $0.012 \pm 0.000052$ & 2 \\
HD 217906 & M2.5II-IIIe & $0.284 \pm 0.033$ & $0.013 \pm 0.00017$ & 2 \\
HD 201251 & K4Ib-IIa & $0.292 \pm 0.034$ & $0.013 \pm 0.000085$ & 2 \\
HD110833 & K3V & $0.306 \pm 0.035$ & $0.014 \pm 0.0001$ & 2 \\
HD39587 & G0VCH+M & $0.308 \pm 0.035$ & $0.014 \pm 0.000091$ & 2 \\
HD134319 & G5V: & $0.432 \pm 0.033$ & $0.019 \pm 0.000069$ & 1 \\
HD41593 & K0V & $0.457 \pm 0.049$ & $0.020 \pm 0.00012$ & 2 \\
HD87884 & K0Ve & $0.458 \pm 0.06$ & $0.020 \pm 0.00018$ & 3 \\
HD 220182 & G9V & $0.467 \pm 0.035$ & $0.021 \pm 0.000051$ & 1 \\
HD47752 & K3.5V & $0.495 \pm 0.074$ & $0.022 \pm 0.00022$ & 4 \\
HD127506 & K3.5V & $0.508 \pm 0.038$ & $0.023 \pm 0.00014$ & 1 \\
HD 122120 & K5V & $0.633 \pm 0.046$ & $0.028 \pm 0.00014$ & 1 \\
HD200560 & K2.5V & $0.638 \pm 0.049$ & $0.028 \pm 0.00045$ & 1 \\
HD  82106 & K3V & $0.638 \pm 0.08$ & $0.028 \pm 0.00017$ & 3 \\
HD 79555 & K4V & $0.651 \pm 0.082$ & $0.029 \pm 0.00024$ & 3 \\
HD 129333 & G1.5V & $0.661 \pm 0.048$ & $0.029 \pm 0.000066$ & 1 \\
HD149957 & K5V & $0.708 \pm 0.057$ & $0.032 \pm 0.0007$ & 1 \\
HD148467 & K6V & $0.727 \pm 0.057$ & $0.032 \pm 0.00062$ & 1 \\
HD 218356 & K1IV(e)+DA1 & $0.834 \pm 0.085$ & $0.037 \pm 0.00022$ & 2 \\
GJ702B & K4V & $0.891 \pm 0.065$ & $0.040 \pm 0.00042$ & 1 \\
HD45088 & K3Vk & $0.963 \pm 0.097$ & $0.043 \pm 0.00028$ & 2 \\
GJ 9781A & K7 & $1.048 \pm 0.075$ & $0.047 \pm 0.00016$ & 1 \\
HD 113827 & K4V & $1.091 \pm 0.078$ & $0.049 \pm 0.00019$ & 1 \\
HD175742 & K0V & $1.339 \pm 0.095$ & $0.060 \pm 0.00034$ & 1 \\
HD151288 & K7V & $1.357 \pm 0.098$ & $0.060 \pm 0.00057$ & 1 \\
HD 88230 & K8V & $1.394 \pm 0.099$ & $0.062 \pm 0.00021$ & 1 \\
HD 266611 & K5V & $1.534 \pm 0.19$ & $0.068 \pm 0.00085$ & 3 \\
\end{tabular}
\end{center}
\end{table}

\added{
\begin{table}
\begin{center}
\caption{$S$-index measurements of HAT-P-11 from the ARC 3.5 m Telescope Echelle Spectrograph (ARCES) at the Apache Point Observatory (APO), calibrated against the Mount Wilson Observatory $S$-index. \label{tab:sind}}
\begin{tabular}{ccc}
JD & $S$ & Uncertainty \\ \hline
2457572.9442 & 0.45 & 0.04 \\
2457575.9589 & 0.43 & 0.04 \\
2457576.8890 & 0.44 & 0.04 \\
2457576.8978 & 0.45 & 0.04 \\
2457576.9065 & 0.45 & 0.04 \\
2457578.9182 & 0.43 & 0.04 \\
2457649.6839 & 0.46 & 0.04 \\
2457649.6964 & 0.46 & 0.04 \\
2457854.8691 & 0.48 & 0.04 \\
2457854.8899 & 0.53 & 0.04 \\
2457854.9056 & 0.55 & 0.05 \\
2457854.9212 & 0.53 & 0.04 \\
2457854.9369 & 0.54 & 0.04 \\
2457854.9527 & 0.54 & 0.04 \\
2457854.9684 & 0.55 & 0.04 \\
2457854.9840 & 0.58 & 0.05 \\
2457916.8112 & 0.53 & 0.04 \\
2457916.8338 & 0.53 & 0.04 \\
2457916.8937 & 0.52 & 0.04 \\
2457916.9163 & 0.53 & 0.04 \\
2457916.9372 & 0.52 & 0.04 \\
2457924.8090 & 0.58 & 0.05 \\
2457924.8260 & 0.58 & 0.05 \\
2458001.6552 & 0.53 & 0.04 \\
2458001.6778 & 0.51 & 0.04 \\
\end{tabular}
\end{center}
\end{table}
}

\end{document}